\documentstyle[twoside,fleqn,bezier,espcrc2]{article}
\newcommand{\tr}[1]{{\rm tr}\,#1}
\newcommand{\ntr}[1]{\,\frac{{\rm tr}}{N}\,#1}
\def\e{{\,\rm e}\,}

\def\be{\begin{equation}}
\def\ee{\end{equation}}
\def\bea{\begin{eqnarray}}
\def\eea{\end{eqnarray}}
\def\LA{\left\langle}
\def\RA{\right\rangle}
\def\RAG{\right\rangle_{\hbox{\footnotesize{Gauss}}}}
\def\RAH{\right\rangle_{\hbox{\footnotesize{Haar}}}}
\newcommand{\rf}[1]{(\ref{#1})}
\newcommand{\eq}[1]{Eq.~(\ref{#1})}

\def\a{\alpha}
\def\b{\beta}
\def\l{\lambda}
\def\ka{\kappa}
\def\om{\omega}
\newcommand{\eps}{\varepsilon}
\newcommand{\non}{\nonumber \\*}
\newcommand{\tV}{\tilde{V}}
\newcommand{\VVp}{{\cal V}^\prime}

\newcommand{\tx}{{x}_-}
\newcommand{\ty}{{x}_+}
\newcommand{\tl}{\lambda}

\newcommand{\re}{\,\hbox{Re}\,}

\newcommand{\ci}{\int_{C_1}\frac{d\omega}{2\pi i}}

\newcommand{\pint}{\int\hspace{-1.15em}\not\hspace{0.6em}}

\newcommand{\ie}{{\it i.e.}\ }
\newcommand{\p}{{\prime}}
\newcommand{\ra}{\rightarrow}
\newcommand{\fr}[2]{{\textstyle {#1 \over #2}}}

\def\fun#1#2{\lower3.6pt\vbox{\baselineskip0pt\lineskip.9pt
\ialign{$\mathsurround=0pt#1\hfil##\hfil$\crcr#2\crcr\sim\crcr}}}

\title{Strings, matrix models, and meanders%
\thanks{Talk at the 29th International Ahrenshoop Symposium
on the Theory of Elementary Particles, Buckow, Germany,
August~29 -- September~2, 1995.}%
\thanks{The research described in this publication
was supported in part by the International Science
Foundation under Grants MF-7000 and MF-7300.}
}

\author{Y. Makeenko\address{
Institute of Theoretical and Experimental Physics,
Russian Federation \\  and
The Niels Bohr Institute, 2100 Copenhagen, Denmark}%
\thanks{E--mail:
makeenko@vxitep.itep.ru \ \ makeenko@nbi.dk } }

\begin{document}

\begin{abstract}
I briefly review the present status of bosonic strings and
discretized random surfaces in $D>1$ which seem to be
in a polymer rather than stringy phase.
As an explicit example of what happens, I consider
the Kazakov--Migdal model with a logarithmic potential
which is exactly solvable for any $D$ (at large $D$
for an arbitrary potential).
I~discuss also the meander problem and report some new results on
its representation via matrix models and the relation to the Kazakov--Migdal
model.  A supersymmetric matrix model is especially useful for describing
the principal meanders.
\end{abstract}

\maketitle

\section{Introduction}

I begin this talk with a
brief review of the present status of the bosonic Polyakov string and
discretized random surfaces in $D>1$ dimensional embedding space.
As an explicit example of what happens, I consider then
the Kazakov--Migdal model with a logarithmic potential
which is exactly solvable for any $D$ (at large $D$
for an arbitrary potential).
I discuss at the end the challenging meander problem  which
is more complicated than the ones solved before by means of
the matrix-model technique but maybe is simpler than the large-$N$ QCD.

\section{Bosonic string in $D>1$}

\subsection{The $D=1$ barrier}

The $D=1$ barrier is associated with the
KPZ (Knizhnik--Polyakov--Zamolodchikov) formula~\cite{KPZ88}
(which was in fact known~\cite{GN84,OW85} before KPZ)
\be
\gamma_{\rm str}=\frac{D-1-\sqrt{(1-D)(25-D)}}{12}
\label{KPZ}
\ee
for the critical index of string susceptibility of the bosonic Polyakov string
in a $D$-dimensional embedding space. Alternatively, it describes
two-dimensional quantum gravity interacting with conformal matter
of the central charge $c=D$.

The right-hand-side (RHS) of \eq{KPZ} is well-defined for $D\leq1$, where
it is associated with topological theories
of gravity which can be described also
by (multi-) matrix models. The RHS becomes complex for $1<D<25$ which is
physically unacceptable. There are two alternatives of how to interpret
this fact:
\begin{itemize}\vspace{-6pt}
\addtolength{\itemsep}{-6pt}
\item
KPZ-approach is not applicable for $D>1$ (say other structures of
conformal theory may become relevant).
\item
KPZ-approach is correct but the interpretation of the result
is not straightforward (say the theory no longer describes a string).
\vspace{-6pt}
\end{itemize}

I discuss below that the second alternative realizes:
the theory is in a branched polymer rather than stringy phase.

\subsection{The DDSW-mechanism}

The mechanism which governs the Polyakov string (= discretized random
surfaces) in $D>1$ was discovered by Das, Dhar, Sengupta and
Wadia~\cite{DDSW90}. They considered the curvature matrix models, where
the propagator is modified as
\be
\LA \Phi_{ij} \Phi_{kl} \RAG \Rightarrow A^{-1}_{il} A^{-1}_{kj}
\label{modprop}
\ee
with the Hermitean $N\times N$ matrix $A$ describing an external field.

The modification~\rf{modprop} of the propagator results in the following
modification of the quadratic part of the potential
\be
\tr \Phi^2 \Rightarrow \tr A\Phi A \Phi .
\ee
These models are solved as $N\ra\infty$ for a logarithmic (Penner) potential
in~\cite{CM92} and for an arbitrary potential in~\cite{KSW95}.

The partition function in the external field $A$
acquires the extra multiplier
\be
Z_{\rm A}\propto\prod_{a={\rm vertex}} \ntr \left( A^{-1} \right)^{\Delta_a},
\ee
where $\Delta_a$ is the coordination number of a given vertex $a$
(which describes intrinsic curvature). Representing the matrix $\Phi$ in the
form $\Phi=\Omega^\dagger \Lambda \Omega$ with diagonal $\Lambda$ and
integrating over the unitary matrix $\Omega$, one gets the action
which involves interaction terms in the form of
the products $\tr \Phi^k \cdots \tr \Phi^l$ with $A$-dependent couplings.

The simplest such a modification of the action reads~\cite{DDSW90}
\be
S=\fr N2 \tr \Phi^2+ \fr {Ng}4 \tr \Phi^4
+ g^\prime \tr \Phi^2 \tr \Phi^2 .
\label{simpl}
\ee
The coupling constant $g^\prime$ describes a new kind of interaction due to
touching of surfaces. These touching diagrams do not vanish as $N\ra\infty$
because of the Weingarten arguments~\cite{Wei80}: the smallness ($N^{-2}$)
of connected correlators is compensated by the fact that the action
$\sim N^2$.
More than one touching of the same surfaces is suppressed as $N^{-2}$.

As was shown by DDSW,
the critical index $\gamma_{\rm str}$ can be expressed
at some value of $g^\prime$ via that ($\gamma_{c<1}$) without
the touching interaction by the formula
\be
\gamma_{\rm str} = \frac{\gamma_{c<1}}{\gamma_{c<1}-1}.
\label{DDSW}
\ee
While only $\gamma_{c<1}=-1/2$ is possible for the quartic
self-interaction~\rf{simpl} with $g^\prime=0$, this formula can be extended
to arbitrary $\gamma_{c<1}$ which is associated with the $c<1$
KPZ-formula~\rf{KPZ}.

\subsection{Trees of baby universes}

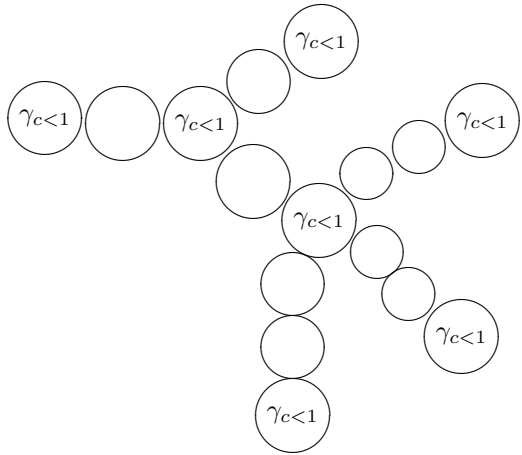
\begin{figure}[tb]
\unitlength=.70mm
\linethickness{0.6pt}
\begin{picture}(92.00,90.50)(-5,0)
\put(49.00,7.00){\circle{14.00}}
\put(49.00,7.00){\makebox(0,0)[cc]{$\gamma_{c<1}$}}
\put(49.00,20.00){\circle{12.00}}
\put(49.00,32.00){\circle{12.17}}
\put(54.00,44.00){\circle{14.00}}
\put(54.00,44.00){\makebox(0,0)[cc]{$\gamma_{c<1}$}}
\put(65.00,38.00){\circle{10.00}}
\put(71.00,30.00){\circle{10.77}}
\put(81.00,22.00){\circle{14.00}}
\put(81.00,22.00){\makebox(0,0)[cc]{$\gamma_{c<1}$}}
\put(63.00,53.00){\circle{10.77}}
\put(73.00,58.00){\circle{10.77}}
\put(85.00,63.00){\circle{14.00}}
\put(85.00,63.00){\makebox(0,0)[cc]{$\gamma_{c<1}$}}
\put(41.50,51.50){\circle{14.00}}
\put(31.50,62.50){\circle{14.00}}
\put(31.50,62.50){\makebox(0,0)[cc]{$\gamma_{c<1}$}}
\put(42.50,70.50){\circle{12.00}}
\put(54.40,78.10){\circle{14.00}}
\put(54.40,78.10){\makebox(0,0)[cc]{$\gamma_{c<1}$}}
\put(16.70,62.50){\circle{14.00}}
\put(1.90,63.50){\circle{14.00}}
\put(1.90,63.50){\makebox(0,0)[cc]{$\gamma_{c<1}$}}
\end{picture}
\caption[x]   {\small
   Trees of 2d baby universes in the $D$-dimensional embedding space.
   2d theory at each surface is critical with
   $\gamma_{c<1}<0$ while \eq{DDSW} with $\gamma_{\rm str}>0$
   in the $D$ dimensions is reached by tuning the coupling of
   the touching interaction between the 2d surfaces.  }
\label{trees}
\end{figure}

Equation~\rf{DDSW} can alternatively be obtained by making a resummation
in the sum over surfaces with touching included. Typical surfaces which
lead to the critical behavior~\rf{DDSW} in $D>1$ are trees constructed
from closed two-dimensional surfaces (baby universes) as is depicted in
Fig.~\ref{trees}.
At each of 2d surfaces, the continuum limit
with $\gamma_{c<1}<0$ is achieved by tuning the cosmological constant
while \eq{DDSW} with $\gamma_{\rm str}>0$ in the $D$-dimensional
embedding space can be reached by tuning the coupling of
the touching interaction.

The case of $\gamma_{c<1}=-1$, which is associated with no critical behavior
at the 2d surface at all, results due to \eq{DDSW} in $\gamma_{\rm str}=1/2$
--- the typical mean-field value for pure branched polymers.

The case of $\gamma_{c<1}=-1/m$, which is associated with
the standard critical behavior of 2d gravity (with matter),
results in $\gamma_{\rm str}=1/(m+1)$ due to polymerization.
It differs from the mean-field value $\gamma_{\rm str}=1/2$
due to effects of 2d gravity. This reminds of how the critical index
$\gamma_{c<1}=-1/3$ appears for the Ising model on a random lattice.

The formula~\rf{DDSW} for the critical behavior of random surfaces
(= Polyakov string) in the $D$-dimensional embedding space
\begin{itemize} \vspace{-6pt}
\addtolength{\itemsep}{-6pt}
\item[1)]
describes numerical data for $c>1$ theories (see~\cite{Amb95} and
references therein),
\item[2)]
can be derived~\cite{Kle95} from the Liouville gravity (by changing
the gravitational dressing from $\exp{(\alpha_+\phi)}$ to
$\exp{(\alpha_-\phi)}$),
\item[3)]
is rigorously proven~\cite{Dur94} assuming locality,
\item[4)]
can be understood at large $D$ (see Subsect.~\ref{l.D.} below).
\vspace{-6pt}
\end{itemize}

Thus the bosonic Polyakov string (= discretized random surfaces) are
in a branched polymer rather than stringy phase for $D>1$. The typical
surfaces are {\it crumpled\/} rather than {\it smooth}.
Such a ground state is stable and has no tachionic excitations.
This picture is
expected for any matrix model describing discretized random surfaces
in $D>1$. I consider in the next Section an explicit example of the
Kazakov--Migdal model~\cite{KM92}.

\section{The Kazakov-Migdal-Penner model~\protect{\cite{Mak95}}}

\subsection{Three equivalent models}

A natural multi-dimensional extension of the matrix chain is
the Kazakov--Migdal (KM) model~\cite{KM92} which
is defined by the partition function
\bea
\lefteqn{Z_{\rm KM}=\int \prod_{\{x,y\}} dU_{xy} \prod_x
 d\phi_x }\non & & \hspace{-.5cm}
\times  \e^{  N \tr{}\left(- \sum_x V(\phi_x)+ c
\sum_{\{x,y\}} \phi_x U^\dagger_{xy} \phi_{y} U_{xy} \right)}.
\label{spartition}
\eea
Here $\phi_x$ and $U_{xy}$ are $N\times N$ Hermitean and unitary
matrices, respectively, with $x$ labeling
lattice sites and $\{xy\}$ labeling the link from the site $x$ to a neighbor
site $y$.

\begin{figure}[tbp]
\unitlength=.50mm
\linethickness{0.6pt}
\begin{picture}(50.0,70.90)(10,60)
\thicklines
\put(40.00,80.00){\line(1,2){20.00}}
\put(60.00,120.00){\line(2,-5){20.00}}
\put(80.00,70.00){\line(-4,1){40.00}}
\put(90.00,90.00){\line(-1,1){30.00}}
\put(80.00,70.00){\line(1,2){10.00}}
\put(60.00,120.00){\line(1,0){20.00}}
\put(90.00,90.00){\line(-1,3){10.00}} 
\thinlines
\put(40.00,80.00){\line(5,1){50.00}}
\put(80.00,70.00){\line(0,1){50.00}}
\put(40.00,80.00){\line(1,1){40.00}}

\put(36.00,79.00){\makebox(0,0)[cc]{1}}
\put(82.00,66.00){\makebox(0,0)[cc]{2}}
\put(58.00,123.00){\makebox(0,0)[cc]{3}}
\put(94.00,90.00){\makebox(0,0)[cc]{4}}
\put(83.00,123.00){\makebox(0,0)[cc]{5}}
\end{picture}
\caption[x]   {\small
   Lattice in the form of a $q$-simplex (depicted for $q=5$).  }
\label{simplex}
\end{figure}
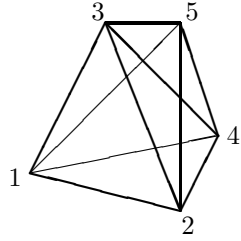

The lattice can be one of the following:
\begin{itemize} \vspace{-6pt}
\addtolength{\itemsep}{-6pt}
\item[i)]
an infinite $D$-dimensional hypercubic lattice~\cite{KM92},
\item[ii)]
a Bethe tree~\cite{Bou93},
\item[iii)]
a $q$-simplex~\cite{Mak95} (see Fig.~\ref{simplex}).
\vspace{-6pt}
\end{itemize}
These three models are equivalent as $N\ra\infty$ at strong coupling
when the coordination numbers
\be
\Delta=2D =q-1
\ee
coincide. In particular, the model on a triangle ($q=3$) is equivalent
to that on a one-dimensional chain ($D=1$). The one-matrix model
is associated with $q=1$ or $D=0$, while the two-matrix model is described by
$q=2$ or $D=1/2$. The gauge field $U_{12}$ can be absorbed in the latter
case by a unitary transformation of $\phi$ so that the standard
Hermitean two-matrix model is recovered.
The proof of equivalence is presented in~\cite{Mak95}.

The integration over the gauge field $U_{xy}$ is over the Haar
measure on SU$(N)$ at each link of the lattice.
The model~\rf{spartition} obviously recovers the standard
open matrix chain if
the lattice is just a one-dimensional sequence of points for which the gauge
field can be absorbed by a unitary transformation of $\phi_x$.

\subsection{Loop equations}

A convenient way of solving the KM model is via the loop
equations which are written for the one-link correlator
\be
G_{\nu\l}= \left\langle
\ntr{\Big(\frac{1}{\nu- \phi_x} U^\dagger_{xy}
\frac{1}{\l- \phi_y} U_{xy} \Big)} \right\rangle.
\label{one-link}
\ee
One gets
\be
G_{\nu\lambda}\stackrel{\nu\ra\infty}{\longrightarrow}
\frac{E_\lambda}{\nu} + \ldots,~
E_\l \equiv \left\langle
\ntr{}\Big( \frac{1}{\l-\phi_x} \Big) \right\rangle
\label{defE}
\ee
at asymptotically large $\nu$.

The loop equation has the same form~\cite{DMS93} as
the one for the two-matrix model
\bea
\lefteqn{ \int_{C_1} \frac{d \om}{2\pi i}
\frac{\VVp(\om)}{(\nu - \om)}\,G_{\om \l}}\non & &=
E_{\nu}\, G_{\nu \l} + c \left( \l G_{\nu \l} - E_\nu \right)
\label{main}
\eea
with the potential
\be
\VVp(\om)\equiv V^\prime(\om)-(\Delta-1) F(\om) ,
\label{defL}
\ee
where $F(\om)$ is determined by the pair correlator of the gauge fields
\be
F\left(\phi_{ij}\right)=c
\frac{\int dU \e^{cN \tr{ \phi U^\dagger
\psi U}} \Big(U^\dagger
\psi U\Big)_{ij}} {\int dU \e^{cN \tr{\phi U^\dagger \psi U}}}.
\label{Lambda}
\ee
The contour $C_1$ in \eq{main} encircles counterclockwise
singularities of the function $G_{\om\l}$ so that the
integration over $\omega$ plays the role of a
projector picking up negative powers of $\nu$.

The $1/\lambda$ term of \eq{main} reads
\be
\int_{C_1}\frac {d\omega}{2\pi i}
\frac{\tilde{V}^\p(\omega)}{(\l-\omega)} E_{\omega} = E_\l^2,
\ee
which coincides with the loop equation for the Hermitean one-matrix
model with the potential
\be
\tilde{V}^\p(\om) =\VVp(\om)-F(\om)= V^\prime(\om)-\Delta F(\om).
\ee
This does not mean that the KM model reduces in general to the one-matrix
model since $\tilde{V}^\p(\om)$ may have singularities outside
of the support of eigenvalues of the master field $\phi_{\rm sp}$ whose
spectral density $\rho(\om)$ is parametrized by $\tilde{V}(\om)$ as
\be
\tilde{V}^\p(\om) =2 \pint dx \frac{\rho(x)}{\om-x}.
\ee

The solution to \eq{main} consists of the following steps.
\begin{itemize} \vspace{-6pt}
\addtolength{\itemsep}{-6pt}
\item[1)]
Given $\VVp(\om)$ find $G_{\nu\l}$ and $E_\l$.
\item[2)]
Calculate $\tilde{V}^\p(\l)=2 \re E_\l$ at the cut.
\item[3)]
Then $V^\p=\Delta \VVp -(1-\Delta)\tilde{V}^\p$ and $F=\VVp-\tilde{V}^\p$
are completely determined.
\vspace{-6pt}
\end{itemize}
It is evident that $V=\tilde{V}$ for the one-matrix model ($\Delta=0$)
and $V={\cal V}$ for the two-matrix model ($\Delta=1$) when the solutions
are known.

\subsection{Two explicit solutions}

The exact solutions of the KM model are known for
\begin{itemize}\vspace{-6pt}
\addtolength{\itemsep}{-6pt}
\item
the Gaussian (quadratic) potential~\cite{Gro92},
\item
the Penner (logarithmic) potential~\cite{Mak93}.
\vspace{-6pt}
\end{itemize}

The quadratic potential can be conveniently parametrized as
\be
V(\phi)=\frac{m_0^2}{2} \phi^2=\frac 12 \left[
\frac ab +(2D-1) c^2 \frac ba \right] \phi^2 ,
\label{quadratic}
\ee
where the parameter $a/b$ is $D$-independent.
The solution is then described by the one-matrix model with the potential
\be
\tilde{V}(\phi)=\frac 12 \left[
\frac ab - c^2\frac ba \right] \phi^2 ,
\ee
which can be obtained substituting $D=0$ in \eq{quadratic}.
The function~\rf{Lambda} reads
\be
F(\phi)=c^2 \frac ba \phi.
\ee

The exactly solvable logarithmic potential reads in the same notations
\bea
\lefteqn{V(\phi)= -\a \ln{(b-\phi)}- (2D-1)
(\a+1)}\non & & \hspace*{.5cm} \times \ln{(a+c\phi)}  + [(2D-1)cb-a] \phi
\label{V}
\eea
and
\be
F(\phi)=c \left[ b-\frac {\a+1}{a+c \phi} \right].
\ee
The Gaussian solution is recovered when
\be
\a=ab,~~a\sim b \ra\infty.
\ee
As is shown in~\cite{Mak93}, the quartic potential is reproduced in the
naive continuum limit for $D<4$.

Introducing new variables
\be
\phi \ra \frac{\phi}{c} +\frac{cb-a}{2c}\,\hbox{I}  ,~~
\b = \frac{a+cb}{2} ,
\label{beta}
\ee
one gets
\bea
\lefteqn{\tilde{V}(\phi)= -\a \ln{(\b-\phi)}} \non
& & \hspace*{.5cm}+ (\a+1) \ln{(\b+\phi)} - 2\b \phi .
\label{htV}
\eea

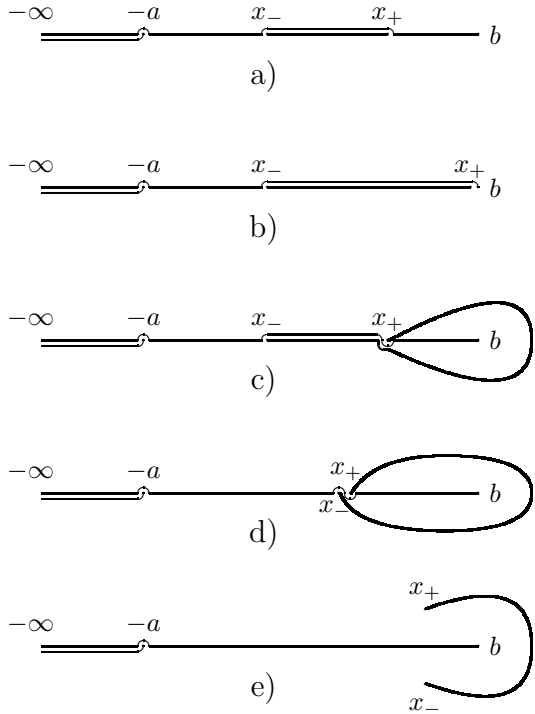
\begin{figure}[tb]
\unitlength=.50mm
\linethickness{0.6pt}
\begin{picture}(150.1,35.00)(-25,80)
\put(104.00,100.00){\circle*{1.00}}
\put(104.00,100.00){\line(-1,0){22.50}}
\put(80.00,100.00){\oval(3.00,3.00)[rt]}
\put(80.00,101.50){\line(-1,0){32.00}}
\put(48.00,100.00){\oval(3.00,3.00)[lt]}
\put(46.50,100.00){\line(-1,0){30.00}}
\put(15.00,100.00){\oval(3.00,3.00)[t]}
\put(15.00,100.00){\circle*{1.00}}
\put(13.50,100.00){\line(-1,0){25.50}}
\put(13.50,100.00){\oval(3.00,3.00)[rb]}
\put(13.50,98.50){\line(-1,0){25.50}}
\thicklines
\put(48.00,100.00){\circle*{1.00}}
\put(48.00,100.00){\line(1,0){32.00}}
\put(80.00,100.00){\circle*{1.00}}
\put(48.00,104.00){\makebox(0,0)[cb]{$x_-$}}
\put(80.00,104.00){\makebox(0,0)[cb]{$x_+$}}
\put(107.00,100.00){\makebox(0,0)[lc]{$b$}}
\put(15.00,104.00){\makebox(0,0)[cb]{$-a$}}
\put(-15.00,104.00){\makebox(0,0)[cb]{$-\infty$}}
\put(47.00,89.00){\makebox(0,0)[cc]{{\large a)}}}
\end{picture}
\begin{picture}(150.1,40.00)(-25,80)
\put(104.00,100.00){\circle*{1.00}}
\put(102.50,100.00){\oval(3.00,3.00)[rt]}
\put(102.50,101.50){\line(-1,0){54.50}}
\put(48.00,100.00){\oval(3.00,3.00)[lt]}
\put(46.50,100.00){\line(-1,0){30.00}}
\put(15.00,100.00){\oval(3.00,3.00)[t]}
\put(15.00,100.00){\circle*{1.00}}
\put(13.50,100.00){\line(-1,0){25.50}}
\put(13.50,100.00){\oval(3.00,3.00)[rb]}
\put(13.50,98.50){\line(-1,0){25.50}}
\thicklines
\put(48.00,100.00){\circle*{1.00}}
\put(48.00,100.00){\line(1,0){54.00}}
\put(102.00,100.00){\circle*{1.00}}
\put(48.00,104.00){\makebox(0,0)[cb]{$x_-$}}
\put(102.00,104.00){\makebox(0,0)[cb]{$x_+$}}
\put(107.00,100.00){\makebox(0,0)[lc]{$b$}}
\put(15.00,104.00){\makebox(0,0)[cb]{$-a$}}
\put(-15.00,104.00){\makebox(0,0)[cb]{$-\infty$}}
\put(47.00,89.00){\makebox(0,0)[cc]{{\large b)}}}
\end{picture}
\begin{picture}(150.1,40.00)(-25,80)
\put(104.00,100.00){\circle*{1.00}}
\put(104.00,100.00){\line(-1,0){22.50}}
\put(80.00,100.00){\oval(3.00,3.00)[b]}
\put(77.00,100.00){\oval(3.00,3.00)[rt]}
\put(77.00,101.50){\line(-1,0){29.00}}
\put(48.00,100.00){\oval(3.00,3.00)[lt]}
\put(46.50,100.00){\line(-1,0){30.00}}
\put(15.00,100.00){\oval(3.00,3.00)[t]}
\put(15.00,100.00){\circle*{1.00}}
\put(13.50,100.00){\line(-1,0){25.50}}
\put(13.50,100.00){\oval(3.00,3.00)[rb]}
\put(13.50,98.50){\line(-1,0){25.50}}
\thicklines
\put(48.00,100.00){\circle*{1.00}}
\put(48.00,100.00){\line(1,0){29.50}}
\put(80.00,100.00){\oval(5.00,5.00)[lb]}
\bezier{256}(80.00,97.50)(118.00,80.00)(118.00,100.00)
\bezier{244}(80.00,100.00)(118.00,120.00)(118.00,100.00)
\put(80.00,100.00){\circle*{1.00}}
\put(48.00,104.00){\makebox(0,0)[cb]{$x_-$}}
\put(80.00,104.00){\makebox(0,0)[cb]{$x_+$}}
\put(107.00,100.00){\makebox(0,0)[lc]{$b$}}
\put(15.00,104.00){\makebox(0,0)[cb]{$-a$}}
\put(-15.00,104.00){\makebox(0,0)[cb]{$-\infty$}}
\put(47.00,89.00){\makebox(0,0)[cc]{{\large c)}}}
\end{picture}
\begin{picture}(150.1,40.00)(-25,80)
\put(104.00,100.00){\circle*{1.00}}
\put(104.00,100.00){\line(-1,0){32.50}}
\put(70.00,100.00){\oval(3.00,3.00)[b]}
\put(67.00,100.00){\oval(3.00,3.00)[t]}
\put(65.50,100.00){\line(-1,0){49.00}}
\put(15.00,100.00){\oval(3.00,3.00)[t]}
\put(15.00,100.00){\circle*{1.00}}
\put(13.50,100.00){\line(-1,0){25.50}}
\put(13.50,100.00){\oval(3.00,3.00)[rb]}
\put(13.50,98.50){\line(-1,0){25.50}}
\thicklines
\put(67.00,100.00){\circle*{1.00}}
\bezier{160}(67.00,100.00)(71.00,90.00)(95.00,90.00)
\bezier{160}(95.00,90.00)(118.00,90.00)(118.00,100.00)
\bezier{155}(95.00,110.00)(118.00,110.00)(118.00,100.00)
\bezier{155}(70.00,100.00)(76.00,110.00)(95.00,110.00)
\put(70.00,100.00){\circle*{1.00}}
\put(66.00,95.00){\makebox(0,0)[cb]{$x_-$}}
\put(69.00,105.00){\makebox(0,0)[cb]{$x_+$}}
\put(107.00,100.00){\makebox(0,0)[lc]{$b$}}
\put(15.00,104.00){\makebox(0,0)[cb]{$-a$}}
\put(-15.00,104.00){\makebox(0,0)[cb]{$-\infty$}}
\put(47.00,89.00){\makebox(0,0)[cc]{{\large d)}}}
\end{picture}
\begin{picture}(150.1,40.00)(-25,80)
\put(104.00,100.00){\circle*{1.00}}
\put(104.00,100.00){\line(-1,0){87.50}}
\put(15.00,100.00){\oval(3.00,3.00)[t]}
\put(15.00,100.00){\circle*{1.00}}
\put(13.50,100.00){\line(-1,0){25.50}}
\put(13.50,100.00){\oval(3.00,3.00)[rb]}
\put(13.50,98.50){\line(-1,0){25.50}}
\thicklines
\put(90.00,90.00){\circle*{1.00}}
\bezier{216}(90.00,90.00)(118.00,80.00)(118.00,100.00)
\bezier{200}(90.00,110.00)(118.00,120.00)(118.00,100.00)
\put(90.00,110.00){\circle*{1.00}}
\put(90.00,86.00){\makebox(0,0)[ct]{$x_-$}}
\put(90.00,114.00){\makebox(0,0)[cb]{$x_+$}}
\put(107.00,100.00){\makebox(0,0)[lc]{$b$}}
\put(15.00,104.00){\makebox(0,0)[cb]{$-a$}}
\put(-15.00,104.00){\makebox(0,0)[cb]{$-\infty$}}
\put(47.00,89.00){\makebox(0,0)[cc]{{\large e)}}}
\end{picture}
\caption[x]
{\small
   Eigenvalue support of the spectral density (the bold
   line) and the branch cuts of the logarithms (the thin lines):
   a) for $\a>0$, b) for $\a\ra+0$, c) for $-1<\a<0$, d)
   for $\a\ra-1$ and e) for $\a<-1$.  }
\label{Fig.1}
\end{figure}

The one-cut solution of the one-matrix model with the potential~\rf{htV}
reads~\cite{Mak95,PW95}
\be
E_\l = \frac {\tilde{V}^\prime(\l)}2
- \sqrt{(\l-\tx)(\l-\ty)} \frac{\b(z+\l)}{\b^2-\tl^2}
\label{one-cut}
\ee
where $z$ is determined by the cubic equation
\be
z^3 - z\left(\b^2 -\a -\fr 12 \right) +\frac{\b}{2} =0
\label{ce}
\ee
and
\be
 {x}_\pm = z - \frac{1}{2\b} \pm
\frac{ \sqrt{\left(\b^2-z^2\right)(4\b z -1)} }{2\b z}
\label{roots}
\ee
are the ends of the cut.
The behavior of the eigen\-value support and branched cuts of the logarithms
are depicted in Fig.~\ref{Fig.1} for various values of $\a$.

\subsection{Critical behavior}

While the only continuum limit of the KM model with the quadratic
potential is possible in $D=1$, the logarithmic potential~\rf{V}
reveals a rich phase structure.
The critical behavior emerges when:
\begin{itemize} \vspace{-6pt}
\addtolength{\itemsep}{-6pt}
\item[i)]
The spectral density ceases to be positive at the interval $[\tx,\ty]$
as for the one-matrix models with a polynomial potential.
\item[ii)]
$\tx$ approaches $-\b$.
\item[iii)]
$\tx$ approaches $\ty$.
\vspace{-6pt}
\end{itemize}

\begin{figure}[t]
\centering
\setlength{\unitlength}{0.240900pt}
\ifx\plotpoint\undefined\newsavebox{\plotpoint}\fi
\sbox{\plotpoint}{\rule[-0.500pt]{1.000pt}{1.000pt}}%
\begin{picture}(900,900)(0,0)
\font\gnuplot=cmr10 at 10pt
\gnuplot
\sbox{\plotpoint}{\rule[-0.500pt]{1.000pt}{1.000pt}}%
\put(220.0,473.0){\rule[-0.500pt]{148.394pt}{1.000pt}}
\put(220.0,68.0){\rule[-0.500pt]{1.000pt}{194.888pt}}
\put(220.0,405.0){\rule[-0.500pt]{4.818pt}{1.000pt}}
\put(198,405){\makebox(0,0)[r]{$-1$}}
\put(816.0,405.0){\rule[-0.500pt]{4.818pt}{1.000pt}}
\put(220.0,877.0){\rule[-0.500pt]{4.818pt}{1.000pt}}
\put(198,877){\makebox(0,0)[r]{$\infty$}}
\put(816.0,877.0){\rule[-0.500pt]{4.818pt}{1.000pt}}
\put(220.0,473.0){\rule[-0.500pt]{4.818pt}{1.000pt}}
\put(198,473){\makebox(0,0)[r]{$0$}}
\put(816.0,473.0){\rule[-0.500pt]{4.818pt}{1.000pt}}
\put(510.0,68.0){\rule[-0.500pt]{1.000pt}{4.818pt}}
\put(510,23){\makebox(0,0){$\sqrt{2}$}}
\put(510.0,857.0){\rule[-0.500pt]{1.000pt}{4.818pt}}
\put(323.0,68.0){\rule[-0.500pt]{1.000pt}{4.818pt}}
\put(323,23){\makebox(0,0){$\frac 12$}}
\put(323.0,857.0){\rule[-0.500pt]{1.000pt}{4.818pt}}
\put(220.0,68.0){\rule[-0.500pt]{1.000pt}{4.818pt}}
\put(220,23){\makebox(0,0){$0$}}
\put(220.0,857.0){\rule[-0.500pt]{1.000pt}{4.818pt}}
\put(836.0,68.0){\rule[-0.500pt]{1.000pt}{4.818pt}}
\put(836,23){\makebox(0,0){$\infty$}}
\put(836.0,857.0){\rule[-0.500pt]{1.000pt}{4.818pt}}
\put(220.0,68.0){\rule[-0.500pt]{148.394pt}{1.000pt}}
\put(836.0,68.0){\rule[-0.500pt]{1.000pt}{194.888pt}}
\put(220.0,877.0){\rule[-0.500pt]{148.394pt}{1.000pt}}
\put(45,472){\makebox(0,0){${\large \alpha}$}}
\put(528,-59){\makebox(0,0)[l]{${\large \beta}$}}
\put(713,473){\makebox(0,0)[l]{\shortstack{{\small Penner} \\ \mbox{} \\
{\small limit}}}}
\put(733,668){\makebox(0,0)[l]{\shortstack{{\small Gauss} \\ \mbox{} \\{\small
limit}}}}
\put(692,810){\makebox(0,0)[l]{\shortstack{{\small cubic} \\ \mbox{} \\{\small
limit}}}}
\put(466,540){\makebox(0,0)[r]{$\alpha=\alpha_c$}}
\put(549,452){\makebox(0,0)[l]{$\alpha=0$}}
\put(549,385){\makebox(0,0)[l]{$\alpha=-1$}}
\put(590,203){\makebox(0,0)[r]{one-cut solution}}
\put(590,742){\makebox(0,0)[r]{two-cut solution}}
\put(220.0,68.0){\rule[-0.500pt]{1.000pt}{194.888pt}}
\put(487,540){\vector(1,0){103}}
\multiput(361.67,506.37)(-0.499,-0.669){142}{\rule{0.120pt}{1.597pt}}
\multiput(361.92,509.69)(-75.000,-97.686){2}{\rule{1.000pt}{0.798pt}}
\put(289,412){\vector(-3,-4){0}}
\put(744,810){\vector(1,0){40}}
\sbox{\plotpoint}{\rule[-0.300pt]{0.600pt}{0.600pt}}%
\put(226,431){\usebox{\plotpoint}}
\multiput(226.00,429.50)(0.821,-0.503){3}{\rule{1.050pt}{0.121pt}}
\multiput(226.00,429.75)(3.821,-4.000){2}{\rule{0.525pt}{0.600pt}}
\multiput(232.00,425.50)(1.009,-0.503){3}{\rule{1.200pt}{0.121pt}}
\multiput(232.00,425.75)(4.509,-4.000){2}{\rule{0.600pt}{0.600pt}}
\put(239,420.25){\rule{1.350pt}{0.600pt}}
\multiput(239.00,421.75)(3.198,-3.000){2}{\rule{0.675pt}{0.600pt}}
\put(245,417.75){\rule{1.445pt}{0.600pt}}
\multiput(245.00,418.75)(3.000,-2.000){2}{\rule{0.723pt}{0.600pt}}
\put(251,415.25){\rule{1.350pt}{0.600pt}}
\multiput(251.00,416.75)(3.198,-3.000){2}{\rule{0.675pt}{0.600pt}}
\put(257,412.75){\rule{1.686pt}{0.600pt}}
\multiput(257.00,413.75)(3.500,-2.000){2}{\rule{0.843pt}{0.600pt}}
\put(264,411.25){\rule{1.445pt}{0.600pt}}
\multiput(264.00,411.75)(3.000,-1.000){2}{\rule{0.723pt}{0.600pt}}
\put(270,409.75){\rule{1.445pt}{0.600pt}}
\multiput(270.00,410.75)(3.000,-2.000){2}{\rule{0.723pt}{0.600pt}}
\put(276,408.25){\rule{1.445pt}{0.600pt}}
\multiput(276.00,408.75)(3.000,-1.000){2}{\rule{0.723pt}{0.600pt}}
\put(282,407.25){\rule{1.445pt}{0.600pt}}
\multiput(282.00,407.75)(3.000,-1.000){2}{\rule{0.723pt}{0.600pt}}
\put(288,406.25){\rule{1.686pt}{0.600pt}}
\multiput(288.00,406.75)(3.500,-1.000){2}{\rule{0.843pt}{0.600pt}}
\put(295,405.25){\rule{1.445pt}{0.600pt}}
\multiput(295.00,405.75)(3.000,-1.000){2}{\rule{0.723pt}{0.600pt}}
\put(307,404.25){\rule{1.445pt}{0.600pt}}
\multiput(307.00,404.75)(3.000,-1.000){2}{\rule{0.723pt}{0.600pt}}
\put(301.0,406.0){\rule[-0.300pt]{1.445pt}{0.600pt}}
\put(332,404.25){\rule{1.445pt}{0.600pt}}
\multiput(332.00,403.75)(3.000,1.000){2}{\rule{0.723pt}{0.600pt}}
\put(313.0,405.0){\rule[-0.300pt]{4.577pt}{0.600pt}}
\put(344,405.25){\rule{1.686pt}{0.600pt}}
\multiput(344.00,404.75)(3.500,1.000){2}{\rule{0.843pt}{0.600pt}}
\put(351,406.25){\rule{1.445pt}{0.600pt}}
\multiput(351.00,405.75)(3.000,1.000){2}{\rule{0.723pt}{0.600pt}}
\put(338.0,406.0){\rule[-0.300pt]{1.445pt}{0.600pt}}
\put(363,407.75){\rule{1.445pt}{0.600pt}}
\multiput(363.00,406.75)(3.000,2.000){2}{\rule{0.723pt}{0.600pt}}
\put(369,409.25){\rule{1.686pt}{0.600pt}}
\multiput(369.00,408.75)(3.500,1.000){2}{\rule{0.843pt}{0.600pt}}
\put(376,410.25){\rule{1.445pt}{0.600pt}}
\multiput(376.00,409.75)(3.000,1.000){2}{\rule{0.723pt}{0.600pt}}
\put(382,411.75){\rule{1.445pt}{0.600pt}}
\multiput(382.00,410.75)(3.000,2.000){2}{\rule{0.723pt}{0.600pt}}
\put(388,413.25){\rule{1.445pt}{0.600pt}}
\multiput(388.00,412.75)(3.000,1.000){2}{\rule{0.723pt}{0.600pt}}
\put(394,414.75){\rule{1.445pt}{0.600pt}}
\multiput(394.00,413.75)(3.000,2.000){2}{\rule{0.723pt}{0.600pt}}
\put(400,416.75){\rule{1.686pt}{0.600pt}}
\multiput(400.00,415.75)(3.500,2.000){2}{\rule{0.843pt}{0.600pt}}
\put(407,418.75){\rule{1.445pt}{0.600pt}}
\multiput(407.00,417.75)(3.000,2.000){2}{\rule{0.723pt}{0.600pt}}
\put(413,421.25){\rule{1.350pt}{0.600pt}}
\multiput(413.00,419.75)(3.198,3.000){2}{\rule{0.675pt}{0.600pt}}
\put(419,423.75){\rule{1.445pt}{0.600pt}}
\multiput(419.00,422.75)(3.000,2.000){2}{\rule{0.723pt}{0.600pt}}
\put(425,425.75){\rule{1.686pt}{0.600pt}}
\multiput(425.00,424.75)(3.500,2.000){2}{\rule{0.843pt}{0.600pt}}
\put(432,428.25){\rule{1.350pt}{0.600pt}}
\multiput(432.00,426.75)(3.198,3.000){2}{\rule{0.675pt}{0.600pt}}
\put(438,431.25){\rule{1.350pt}{0.600pt}}
\multiput(438.00,429.75)(3.198,3.000){2}{\rule{0.675pt}{0.600pt}}
\put(444,434.25){\rule{1.350pt}{0.600pt}}
\multiput(444.00,432.75)(3.198,3.000){2}{\rule{0.675pt}{0.600pt}}
\put(450,437.25){\rule{1.350pt}{0.600pt}}
\multiput(450.00,435.75)(3.198,3.000){2}{\rule{0.675pt}{0.600pt}}
\put(456,440.25){\rule{1.550pt}{0.600pt}}
\multiput(456.00,438.75)(3.783,3.000){2}{\rule{0.775pt}{0.600pt}}
\multiput(463.00,443.99)(0.821,0.503){3}{\rule{1.050pt}{0.121pt}}
\multiput(463.00,441.75)(3.821,4.000){2}{\rule{0.525pt}{0.600pt}}
\put(469,447.25){\rule{1.350pt}{0.600pt}}
\multiput(469.00,445.75)(3.198,3.000){2}{\rule{0.675pt}{0.600pt}}
\multiput(475.00,450.99)(0.821,0.503){3}{\rule{1.050pt}{0.121pt}}
\multiput(475.00,448.75)(3.821,4.000){2}{\rule{0.525pt}{0.600pt}}
\multiput(481.00,454.99)(1.009,0.503){3}{\rule{1.200pt}{0.121pt}}
\multiput(481.00,452.75)(4.509,4.000){2}{\rule{0.600pt}{0.600pt}}
\put(488,458.25){\rule{1.350pt}{0.600pt}}
\multiput(488.00,456.75)(3.198,3.000){2}{\rule{0.675pt}{0.600pt}}
\multiput(494.00,461.99)(0.821,0.503){3}{\rule{1.050pt}{0.121pt}}
\multiput(494.00,459.75)(3.821,4.000){2}{\rule{0.525pt}{0.600pt}}
\multiput(500.00,465.99)(0.597,0.502){5}{\rule{0.870pt}{0.121pt}}
\multiput(500.00,463.75)(4.194,5.000){2}{\rule{0.435pt}{0.600pt}}
\multiput(506.00,470.99)(0.821,0.503){3}{\rule{1.050pt}{0.121pt}}
\multiput(506.00,468.75)(3.821,4.000){2}{\rule{0.525pt}{0.600pt}}
\multiput(512.00,474.99)(1.009,0.503){3}{\rule{1.200pt}{0.121pt}}
\multiput(512.00,472.75)(4.509,4.000){2}{\rule{0.600pt}{0.600pt}}
\multiput(519.00,478.99)(0.597,0.502){5}{\rule{0.870pt}{0.121pt}}
\multiput(519.00,476.75)(4.194,5.000){2}{\rule{0.435pt}{0.600pt}}
\multiput(525.00,483.99)(0.597,0.502){5}{\rule{0.870pt}{0.121pt}}
\multiput(525.00,481.75)(4.194,5.000){2}{\rule{0.435pt}{0.600pt}}
\multiput(531.00,488.99)(0.597,0.502){5}{\rule{0.870pt}{0.121pt}}
\multiput(531.00,486.75)(4.194,5.000){2}{\rule{0.435pt}{0.600pt}}
\multiput(537.00,493.99)(0.723,0.502){5}{\rule{0.990pt}{0.121pt}}
\multiput(537.00,491.75)(4.945,5.000){2}{\rule{0.495pt}{0.600pt}}
\multiput(544.00,498.99)(0.597,0.502){5}{\rule{0.870pt}{0.121pt}}
\multiput(544.00,496.75)(4.194,5.000){2}{\rule{0.435pt}{0.600pt}}
\multiput(550.00,503.99)(0.597,0.502){5}{\rule{0.870pt}{0.121pt}}
\multiput(550.00,501.75)(4.194,5.000){2}{\rule{0.435pt}{0.600pt}}
\multiput(556.00,508.99)(0.597,0.502){5}{\rule{0.870pt}{0.121pt}}
\multiput(556.00,506.75)(4.194,5.000){2}{\rule{0.435pt}{0.600pt}}
\multiput(562.00,513.99)(0.481,0.501){7}{\rule{0.750pt}{0.121pt}}
\multiput(562.00,511.75)(4.443,6.000){2}{\rule{0.375pt}{0.600pt}}
\multiput(568.00,519.99)(0.723,0.502){5}{\rule{0.990pt}{0.121pt}}
\multiput(568.00,517.75)(4.945,5.000){2}{\rule{0.495pt}{0.600pt}}
\multiput(575.00,524.99)(0.481,0.501){7}{\rule{0.750pt}{0.121pt}}
\multiput(575.00,522.75)(4.443,6.000){2}{\rule{0.375pt}{0.600pt}}
\multiput(581.00,530.99)(0.481,0.501){7}{\rule{0.750pt}{0.121pt}}
\multiput(581.00,528.75)(4.443,6.000){2}{\rule{0.375pt}{0.600pt}}
\multiput(587.00,536.99)(0.481,0.501){7}{\rule{0.750pt}{0.121pt}}
\multiput(587.00,534.75)(4.443,6.000){2}{\rule{0.375pt}{0.600pt}}
\multiput(593.00,542.99)(0.579,0.501){7}{\rule{0.850pt}{0.121pt}}
\multiput(593.00,540.75)(5.236,6.000){2}{\rule{0.425pt}{0.600pt}}
\multiput(600.99,548.00)(0.501,0.579){7}{\rule{0.121pt}{0.850pt}}
\multiput(598.75,548.00)(6.000,5.236){2}{\rule{0.600pt}{0.425pt}}
\multiput(606.00,555.99)(0.481,0.501){7}{\rule{0.750pt}{0.121pt}}
\multiput(606.00,553.75)(4.443,6.000){2}{\rule{0.375pt}{0.600pt}}
\multiput(612.99,561.00)(0.501,0.579){7}{\rule{0.121pt}{0.850pt}}
\multiput(610.75,561.00)(6.000,5.236){2}{\rule{0.600pt}{0.425pt}}
\multiput(618.00,568.99)(0.481,0.501){7}{\rule{0.750pt}{0.121pt}}
\multiput(618.00,566.75)(4.443,6.000){2}{\rule{0.375pt}{0.600pt}}
\multiput(624.00,574.99)(0.486,0.501){9}{\rule{0.750pt}{0.121pt}}
\multiput(624.00,572.75)(5.443,7.000){2}{\rule{0.375pt}{0.600pt}}
\multiput(631.99,581.00)(0.501,0.579){7}{\rule{0.121pt}{0.850pt}}
\multiput(629.75,581.00)(6.000,5.236){2}{\rule{0.600pt}{0.425pt}}
\multiput(637.99,588.00)(0.501,0.579){7}{\rule{0.121pt}{0.850pt}}
\multiput(635.75,588.00)(6.000,5.236){2}{\rule{0.600pt}{0.425pt}}
\multiput(643.99,595.00)(0.501,0.579){7}{\rule{0.121pt}{0.850pt}}
\multiput(641.75,595.00)(6.000,5.236){2}{\rule{0.600pt}{0.425pt}}
\multiput(649.99,602.00)(0.501,0.566){9}{\rule{0.121pt}{0.836pt}}
\multiput(647.75,602.00)(7.000,6.265){2}{\rule{0.600pt}{0.418pt}}
\multiput(656.99,610.00)(0.501,0.579){7}{\rule{0.121pt}{0.850pt}}
\multiput(654.75,610.00)(6.000,5.236){2}{\rule{0.600pt}{0.425pt}}
\multiput(662.99,617.00)(0.501,0.676){7}{\rule{0.121pt}{0.950pt}}
\multiput(660.75,617.00)(6.000,6.028){2}{\rule{0.600pt}{0.475pt}}
\multiput(668.99,625.00)(0.501,0.579){7}{\rule{0.121pt}{0.850pt}}
\multiput(666.75,625.00)(6.000,5.236){2}{\rule{0.600pt}{0.425pt}}
\multiput(674.99,632.00)(0.501,0.676){7}{\rule{0.121pt}{0.950pt}}
\multiput(672.75,632.00)(6.000,6.028){2}{\rule{0.600pt}{0.475pt}}
\multiput(680.99,640.00)(0.501,0.566){9}{\rule{0.121pt}{0.836pt}}
\multiput(678.75,640.00)(7.000,6.265){2}{\rule{0.600pt}{0.418pt}}
\multiput(687.99,648.00)(0.501,0.676){7}{\rule{0.121pt}{0.950pt}}
\multiput(685.75,648.00)(6.000,6.028){2}{\rule{0.600pt}{0.475pt}}
\multiput(693.99,656.00)(0.501,0.774){7}{\rule{0.121pt}{1.050pt}}
\multiput(691.75,656.00)(6.000,6.821){2}{\rule{0.600pt}{0.525pt}}
\multiput(699.99,665.00)(0.501,0.676){7}{\rule{0.121pt}{0.950pt}}
\multiput(697.75,665.00)(6.000,6.028){2}{\rule{0.600pt}{0.475pt}}
\multiput(705.99,673.00)(0.501,0.647){9}{\rule{0.121pt}{0.921pt}}
\multiput(703.75,673.00)(7.000,7.088){2}{\rule{0.600pt}{0.461pt}}
\multiput(712.99,682.00)(0.501,0.676){7}{\rule{0.121pt}{0.950pt}}
\multiput(710.75,682.00)(6.000,6.028){2}{\rule{0.600pt}{0.475pt}}
\multiput(718.99,690.00)(0.501,0.774){7}{\rule{0.121pt}{1.050pt}}
\multiput(716.75,690.00)(6.000,6.821){2}{\rule{0.600pt}{0.525pt}}
\multiput(724.99,699.00)(0.501,0.774){7}{\rule{0.121pt}{1.050pt}}
\multiput(722.75,699.00)(6.000,6.821){2}{\rule{0.600pt}{0.525pt}}
\multiput(730.99,708.00)(0.501,0.774){7}{\rule{0.121pt}{1.050pt}}
\multiput(728.75,708.00)(6.000,6.821){2}{\rule{0.600pt}{0.525pt}}
\multiput(736.99,717.00)(0.501,0.647){9}{\rule{0.121pt}{0.921pt}}
\multiput(734.75,717.00)(7.000,7.088){2}{\rule{0.600pt}{0.461pt}}
\multiput(743.99,726.00)(0.501,0.774){7}{\rule{0.121pt}{1.050pt}}
\multiput(741.75,726.00)(6.000,6.821){2}{\rule{0.600pt}{0.525pt}}
\multiput(749.99,735.00)(0.501,0.871){7}{\rule{0.121pt}{1.150pt}}
\multiput(747.75,735.00)(6.000,7.613){2}{\rule{0.600pt}{0.575pt}}
\multiput(755.99,745.00)(0.501,0.774){7}{\rule{0.121pt}{1.050pt}}
\multiput(753.75,745.00)(6.000,6.821){2}{\rule{0.600pt}{0.525pt}}
\multiput(761.99,754.00)(0.501,0.727){9}{\rule{0.121pt}{1.007pt}}
\multiput(759.75,754.00)(7.000,7.910){2}{\rule{0.600pt}{0.504pt}}
\multiput(768.99,764.00)(0.501,0.871){7}{\rule{0.121pt}{1.150pt}}
\multiput(766.75,764.00)(6.000,7.613){2}{\rule{0.600pt}{0.575pt}}
\multiput(774.99,774.00)(0.501,0.871){7}{\rule{0.121pt}{1.150pt}}
\multiput(772.75,774.00)(6.000,7.613){2}{\rule{0.600pt}{0.575pt}}
\multiput(780.99,784.00)(0.501,0.871){7}{\rule{0.121pt}{1.150pt}}
\multiput(778.75,784.00)(6.000,7.613){2}{\rule{0.600pt}{0.575pt}}
\multiput(786.99,794.00)(0.501,0.871){7}{\rule{0.121pt}{1.150pt}}
\multiput(784.75,794.00)(6.000,7.613){2}{\rule{0.600pt}{0.575pt}}
\multiput(792.99,804.00)(0.501,0.727){9}{\rule{0.121pt}{1.007pt}}
\multiput(790.75,804.00)(7.000,7.910){2}{\rule{0.600pt}{0.504pt}}
\multiput(799.99,814.00)(0.501,0.969){7}{\rule{0.121pt}{1.250pt}}
\multiput(797.75,814.00)(6.000,8.406){2}{\rule{0.600pt}{0.625pt}}
\multiput(805.99,825.00)(0.501,0.871){7}{\rule{0.121pt}{1.150pt}}
\multiput(803.75,825.00)(6.000,7.613){2}{\rule{0.600pt}{0.575pt}}
\multiput(811.99,835.00)(0.501,0.969){7}{\rule{0.121pt}{1.250pt}}
\multiput(809.75,835.00)(6.000,8.406){2}{\rule{0.600pt}{0.625pt}}
\multiput(817.99,846.00)(0.501,0.808){9}{\rule{0.121pt}{1.093pt}}
\multiput(815.75,846.00)(7.000,8.732){2}{\rule{0.600pt}{0.546pt}}
\multiput(824.99,857.00)(0.501,0.969){7}{\rule{0.121pt}{1.250pt}}
\multiput(822.75,857.00)(6.000,8.406){2}{\rule{0.600pt}{0.625pt}}
\multiput(830.99,868.00)(0.502,0.974){5}{\rule{0.121pt}{1.230pt}}
\multiput(828.75,868.00)(5.000,6.447){2}{\rule{0.600pt}{0.615pt}}
\put(357.0,408.0){\rule[-0.300pt]{1.445pt}{0.600pt}}
\sbox{\plotpoint}{\rule[-0.250pt]{0.500pt}{0.500pt}}%
\put(662,477){\usebox{\plotpoint}}
\multiput(662,477)(4.145,11.743){2}{\usebox{\plotpoint}}
\put(671.82,499.72){\usebox{\plotpoint}}
\multiput(674,503)(8.806,8.806){0}{\usebox{\plotpoint}}
\put(680.02,509.02){\usebox{\plotpoint}}
\put(689.51,517.09){\usebox{\plotpoint}}
\multiput(693,520)(10.362,6.908){0}{\usebox{\plotpoint}}
\put(699.53,524.45){\usebox{\plotpoint}}
\put(709.91,531.10){\usebox{\plotpoint}}
\multiput(712,532)(10.362,6.908){0}{\usebox{\plotpoint}}
\put(720.65,537.33){\usebox{\plotpoint}}
\multiput(724,539)(10.362,6.908){0}{\usebox{\plotpoint}}
\put(731.34,543.67){\usebox{\plotpoint}}
\put(742.66,548.85){\usebox{\plotpoint}}
\multiput(743,549)(11.139,5.569){0}{\usebox{\plotpoint}}
\put(754.10,553.70){\usebox{\plotpoint}}
\multiput(755,554)(11.139,5.569){0}{\usebox{\plotpoint}}
\put(765.41,558.89){\usebox{\plotpoint}}
\multiput(768,560)(11.814,3.938){0}{\usebox{\plotpoint}}
\put(776.96,563.48){\usebox{\plotpoint}}
\multiput(780,565)(11.814,3.938){0}{\usebox{\plotpoint}}
\put(788.59,567.86){\usebox{\plotpoint}}
\multiput(792,569)(11.446,4.906){0}{\usebox{\plotpoint}}
\put(800.18,572.39){\usebox{\plotpoint}}
\multiput(805,574)(11.814,3.938){0}{\usebox{\plotpoint}}
\put(811.99,576.33){\usebox{\plotpoint}}
\put(823.90,579.97){\usebox{\plotpoint}}
\multiput(824,580)(11.814,3.938){0}{\usebox{\plotpoint}}
\put(835.71,583.90){\usebox{\plotpoint}}
\put(836,584){\usebox{\plotpoint}}
\put(662,468){\usebox{\plotpoint}}
\multiput(662,468)(4.145,-11.743){2}{\usebox{\plotpoint}}
\put(671.82,445.28){\usebox{\plotpoint}}
\multiput(674,442)(8.806,-8.806){0}{\usebox{\plotpoint}}
\put(680.02,435.98){\usebox{\plotpoint}}
\put(689.51,427.91){\usebox{\plotpoint}}
\multiput(693,425)(10.362,-6.908){0}{\usebox{\plotpoint}}
\put(699.53,420.55){\usebox{\plotpoint}}
\put(709.91,413.90){\usebox{\plotpoint}}
\multiput(712,413)(10.362,-6.908){0}{\usebox{\plotpoint}}
\put(720.65,407.67){\usebox{\plotpoint}}
\multiput(724,406)(10.362,-6.908){0}{\usebox{\plotpoint}}
\put(731.34,401.33){\usebox{\plotpoint}}
\put(742.66,396.15){\usebox{\plotpoint}}
\multiput(743,396)(11.139,-5.569){0}{\usebox{\plotpoint}}
\put(754.10,391.30){\usebox{\plotpoint}}
\multiput(755,391)(11.139,-5.569){0}{\usebox{\plotpoint}}
\put(765.41,386.11){\usebox{\plotpoint}}
\multiput(768,385)(11.814,-3.938){0}{\usebox{\plotpoint}}
\put(776.96,381.52){\usebox{\plotpoint}}
\multiput(780,380)(11.814,-3.938){0}{\usebox{\plotpoint}}
\put(788.59,377.14){\usebox{\plotpoint}}
\multiput(792,376)(11.446,-4.906){0}{\usebox{\plotpoint}}
\put(800.18,372.61){\usebox{\plotpoint}}
\multiput(805,371)(11.814,-3.938){0}{\usebox{\plotpoint}}
\put(811.99,368.67){\usebox{\plotpoint}}
\put(823.90,365.03){\usebox{\plotpoint}}
\multiput(824,365)(11.814,-3.938){0}{\usebox{\plotpoint}}
\put(835.71,361.10){\usebox{\plotpoint}}
\put(836,361){\usebox{\plotpoint}}
\put(662,625){\usebox{\plotpoint}}
\multiput(662,625)(2.047,12.284){3}{\usebox{\plotpoint}}
\multiput(668,661)(3.277,12.014){2}{\usebox{\plotpoint}}
\multiput(674,683)(3.750,11.875){2}{\usebox{\plotpoint}}
\put(683.03,709.35){\usebox{\plotpoint}}
\multiput(687,719)(4.145,11.743){2}{\usebox{\plotpoint}}
\put(696.30,744.26){\usebox{\plotpoint}}
\put(700.93,755.82){\usebox{\plotpoint}}
\multiput(705,766)(5.266,11.285){2}{\usebox{\plotpoint}}
\put(715.66,790.14){\usebox{\plotpoint}}
\put(720.28,801.70){\usebox{\plotpoint}}
\multiput(724,811)(4.906,11.446){2}{\usebox{\plotpoint}}
\put(734.77,836.14){\usebox{\plotpoint}}
\put(740.18,847.35){\usebox{\plotpoint}}
\put(745.42,858.64){\usebox{\plotpoint}}
\put(750.25,870.12){\usebox{\plotpoint}}
\put(753,877){\usebox{\plotpoint}}
\put(662,609){\usebox{\plotpoint}}
\multiput(662,609)(3.143,-12.050){2}{\usebox{\plotpoint}}
\put(668.74,585.14){\usebox{\plotpoint}}
\put(677.64,576.58){\usebox{\plotpoint}}
\multiput(680,575)(11.974,-3.421){0}{\usebox{\plotpoint}}
\put(689.30,572.62){\usebox{\plotpoint}}
\multiput(693,572)(12.453,0.000){0}{\usebox{\plotpoint}}
\put(701.67,572.44){\usebox{\plotpoint}}
\multiput(705,573)(11.974,3.421){0}{\usebox{\plotpoint}}
\put(713.70,575.57){\usebox{\plotpoint}}
\multiput(718,577)(11.139,5.569){0}{\usebox{\plotpoint}}
\put(725.09,580.54){\usebox{\plotpoint}}
\multiput(730,583)(11.139,5.569){0}{\usebox{\plotpoint}}
\put(736.22,586.12){\usebox{\plotpoint}}
\put(746.86,592.58){\usebox{\plotpoint}}
\multiput(749,594)(9.567,7.972){0}{\usebox{\plotpoint}}
\put(756.59,600.33){\usebox{\plotpoint}}
\put(766.47,607.90){\usebox{\plotpoint}}
\multiput(768,609)(9.567,7.972){0}{\usebox{\plotpoint}}
\put(775.95,615.95){\usebox{\plotpoint}}
\put(785.17,624.31){\usebox{\plotpoint}}
\multiput(786,625)(8.806,8.806){0}{\usebox{\plotpoint}}
\put(794.04,633.04){\usebox{\plotpoint}}
\put(802.84,641.84){\usebox{\plotpoint}}
\multiput(805,644)(8.105,9.455){0}{\usebox{\plotpoint}}
\put(811.12,651.14){\usebox{\plotpoint}}
\put(819.42,660.42){\usebox{\plotpoint}}
\put(827.89,669.54){\usebox{\plotpoint}}
\put(835.52,679.37){\usebox{\plotpoint}}
\put(836,680){\usebox{\plotpoint}}
\put(693,659){\usebox{\plotpoint}}
\multiput(693,659)(4.906,11.446){2}{\usebox{\plotpoint}}
\put(703.33,681.65){\usebox{\plotpoint}}
\put(709.68,692.35){\usebox{\plotpoint}}
\put(715.89,703.13){\usebox{\plotpoint}}
\put(722.14,713.90){\usebox{\plotpoint}}
\put(728.23,724.76){\usebox{\plotpoint}}
\put(734.20,735.69){\usebox{\plotpoint}}
\put(740.98,746.12){\usebox{\plotpoint}}
\put(747.28,756.84){\usebox{\plotpoint}}
\put(753.24,767.78){\usebox{\plotpoint}}
\put(759.21,778.71){\usebox{\plotpoint}}
\put(765.99,789.13){\usebox{\plotpoint}}
\put(772.29,799.86){\usebox{\plotpoint}}
\put(778.25,810.79){\usebox{\plotpoint}}
\put(783.93,821.87){\usebox{\plotpoint}}
\put(789.75,832.88){\usebox{\plotpoint}}
\put(796.17,843.55){\usebox{\plotpoint}}
\put(802.44,854.30){\usebox{\plotpoint}}
\put(808.17,865.35){\usebox{\plotpoint}}
\put(813.74,876.49){\usebox{\plotpoint}}
\put(814,877){\usebox{\plotpoint}}
\put(693,654){\usebox{\plotpoint}}
\put(693.00,654.00){\usebox{\plotpoint}}
\put(703.71,659.92){\usebox{\plotpoint}}
\multiput(705,661)(9.455,8.105){0}{\usebox{\plotpoint}}
\put(713.01,668.18){\usebox{\plotpoint}}
\put(721.38,677.38){\usebox{\plotpoint}}
\put(729.25,687.00){\usebox{\plotpoint}}
\multiput(730,688)(8.105,9.455){0}{\usebox{\plotpoint}}
\put(737.31,696.50){\usebox{\plotpoint}}
\put(745.48,705.89){\usebox{\plotpoint}}
\put(752.91,715.86){\usebox{\plotpoint}}
\put(760.21,725.95){\usebox{\plotpoint}}
\multiput(761,727)(8.201,9.372){0}{\usebox{\plotpoint}}
\put(768.28,735.42){\usebox{\plotpoint}}
\put(775.19,745.78){\usebox{\plotpoint}}
\put(782.10,756.14){\usebox{\plotpoint}}
\put(789.00,766.51){\usebox{\plotpoint}}
\put(796.33,776.57){\usebox{\plotpoint}}
\put(803.17,786.95){\usebox{\plotpoint}}
\put(809.93,797.40){\usebox{\plotpoint}}
\put(816.42,808.03){\usebox{\plotpoint}}
\put(823.49,818.28){\usebox{\plotpoint}}
\put(829.95,828.92){\usebox{\plotpoint}}
\multiput(830,829)(6.407,10.679){0}{\usebox{\plotpoint}}
\put(836,839){\usebox{\plotpoint}}
\sbox{\plotpoint}{\rule[-0.500pt]{1.000pt}{1.000pt}}%
\put(220,405){\usebox{\plotpoint}}
\put(220.0,405.0){\rule[-0.500pt]{148.394pt}{1.000pt}}
\sbox{\plotpoint}{\rule[-0.175pt]{0.350pt}{0.350pt}}%
\put(323,405){\raisebox{-.8pt}{\makebox(0,0){$\Diamond$}}}
\put(510,473){\raisebox{-.8pt}{\makebox(0,0){$\Diamond$}}}
\end{picture}
\caption[x]   {\small
   Phase diagram of the KM model with the logarithmic
   potential~\rf{V}. The bold line which starts at
   $\b=0$, $\a=-1/2$ corresponds to
   \eq{discriminant}. The one-cut solution realizes for $\a<\a_c$.
   The critical lines $\a=\a_c$ and $\a=-1$
   correspond to $\gamma_{str}=-1/2$ and $\gamma_{str}=0$, respectively,
   while the tricritical point $\b=1/2$, $\a=-1$ is associated with a
   phase transition of the Kosterlitz--Thouless type.
   }
\label{phases}
\end{figure}
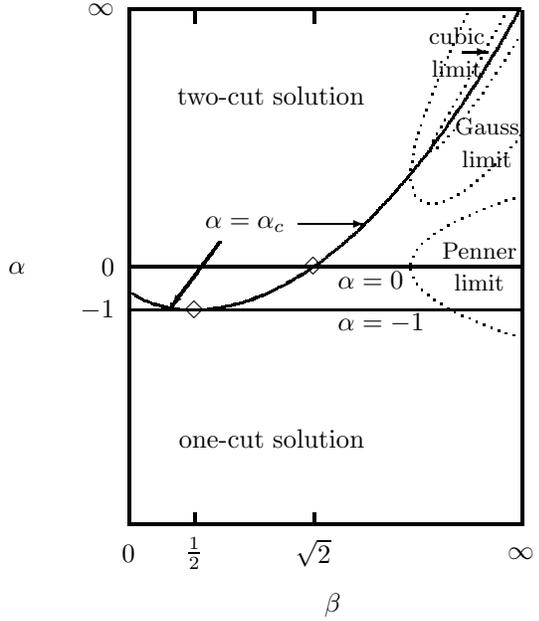

The critical behavior the type $i)$ occurs along the line
\be
 \a_c=\b^2 -3 \left(\frac{\b}{4}\right)^{\frac 23} -\fr 12 ,
\label{discriminant}
\ee
where $\tx = -z$.
At this line the discriminant of the cubic equation~\rf{ce}
vanishes and the one-cut solution is not applicable for $\a>\a_c$.
The critical behavior of the types  $ii)$ and $iii)$
occurs for $\a=0$ and $\a=-1$ respectively.
These critical lines are depicted in Fig.~\ref{phases}.
The critical behavior agrees with that for the one-matrix model with
the cubic and Penner potentials which can be obtained from
of the potential~\rf{htV} in the limits
\be
\a\sim\b^2\ra \infty,~(\b^2-\a)\sim \beta^{2/3}~~\hbox{{\small (cubic limit)}}
\ee
and
\be
\b\ra\infty,~\a\sim 1~~~\hbox{{\small (Penner limit)}},
\ee
respectively.

The character of the critical behavior can be find out by calculating the
susceptibility
\be
\chi \equiv - \frac{1}{N^2\; \hbox{Vol.}} \frac{d^2}{d\a^2} \ln{Z_{\rm KM}}
\ee
where\ \/Vol.\ \/stands for the volume of the system =
the number of sites of the lattice.
The result in genus zero reads explicitly
\bea
\lefteqn{\chi_0 = (D-1) \ln{}\left\{ \frac 14
\left( \sqrt[4]{\frac{(\b+x_-)(\b-x_+)}{(\b+x_+)(\b-x_-)}}
\right. \right.} \non & &~~~~~~~\left.\left.
\pm \sqrt[4]{\frac{(\b+x_+)(\b-x_-)}{(\b+x_-)(\b-x_+)}} \right)^2\right\}
\nonumber \\
& &\hspace*{-.5cm}+D \ln{\left\{ \left( \sqrt[4]{\frac{(\b+x_-)}{(\b+x_+)}} +
\sqrt[4]{\frac{(\b+x_+)}{(\b+x_-)}}
\right)^2\right\}} \non & & \hspace*{-.5cm} - D \ln{\left\{ \left(
\sqrt[4]{\frac{(\b-x_+)}{(\b-x_-)}} \pm
\sqrt[4]{\frac{(\b-x_-)}{(\b-x_+)}}
\right)^2\right\}}.
\label{susc}
\eea
where the positive sign in $\pm$ corresponds to $\a>0$ while the
minus sign should be substituted for $\a<0$.

Having the explicit formula~\rf{susc} for $\chi_0$, we can find out which
$\gamma_{str}$ is associated with each type of
the critical behavior. Near the line~\rf{discriminant} where $\chi_0$
is not singular and equals some value $\chi_0^c$, one gets
\be
\chi_0-\chi_0^c \sim (x_--x_-^c)~~~\hbox{for } \a\approx\a_c.
\label{4.14}
\ee
Since $(x_--x_-^c)\sim (\a_c-\a)^{1/2}$,
one obtains $\gamma_{str}=-1/2$ near the critical line~\rf{discriminant}.

On the contrary, $\chi_0$ given by \eq{susc} has
logarithmic singularities for $\a=0$ and $\a=-1$:
\be
\chi_0 \approx \frac 12 \ln{(\b-x_+)} ~~~\hbox{for } \a\approx 0
\label{sing2}
\ee
and
\be
\chi_0 \approx - \log{(x_--x_+)^2} ~~~\hbox{for } \a\approx-1.
\label{sing1}
\ee
Near the critical lines $\alpha=0$ and $\alpha=-1$,
one gets $\gamma_{str}=0$.
While~\rf{sing1} is positive, \rf{sing2} is negative.
For this reason we exclude
the critical line $\alpha=0$ from the consideration.

\subsection{Continuum limits}
\subsubsection{$\gamma_{srt}=-1/2$}

The continuum theory near $\alpha=\alpha_c$ given by \eq{discriminant}
reminds 2d gravity and can be obtained expanding near the edge singularity:
\be
z=\left( \frac{\b}{4} \right)^{\frac13}
 + \eps \sqrt{\Lambda} ,~\lambda=\left( \frac{\b}{4} \right)^{\frac13}
 + \eps \xi ,
 \label{6.1}
\ee
where $\eps\ra0$, $\Lambda$ is the cosmological constant, and
$\xi$ is the continuum momentum variable.

The continuum spectral density is determined by \eq{one-cut} to be
\be
\rho_c (\xi) \propto \frac 1\pi \left( \xi + \sqrt{\Lambda} \right)
\sqrt{\xi - 2 \sqrt{\Lambda}}
\label{rhocont}
\ee
and describes
all continuum correlators of the trace of powers of the (renormalized) field
$\phi_x$ at some point $x$, which is the standard set of observables of $2d$
gravity.

The critical behavior of matter is described
by observables associated with
extended objects --- the open-loop averages
\bea
\lefteqn{G_{\nu\l}(C_{xy})=} \non & & \left\langle
\ntr{\Big(\frac{1}{\nu- \phi_x} U^\dagger_{xy}
\frac{1}{\l- \phi_y}
U_{xy} \Big)} \right\rangle,
\label{KMG}
\eea
where $C_{xy}$ goes from $x$ to $y$ along some path
and the average is w.r.t.\ the same measure as in~\rf{spartition}.
They depend at large $N$
only on the algebraic length $L(C_{xy})$ of
the contour $C_{xy}$.

The double discontinuity
\bea
\lefteqn{C(\nu,\l;L) } \non & &
\equiv \frac{1}{\pi^2 \rho(\nu) \rho(\l)}\,\hbox{Disc}_\nu\,
\hbox{Disc}_\l\, G_{\nu\l}(C_{xy}) .
\label{defCL}
\eea
across the cut determines
$C(\nu,\l;1)$ --- the one-link Itzykson--Zuber correlator
of the gauge fields which reads
\be
C(\nu,\l;1) \propto \frac {1}{\eps^2(\xi_\nu-\xi_\l)^2}
\label{Cone-link}
\ee
as $\eps\ra0$. It is quite similar to the Gaussian one in the naive
continuum limit at $D=1$.

The RHS of \eq{Cone-link} describes $C(\nu,\l;L)$ for $L\ll 1/\sqrt{\eps}$
while a nontrivial continuum limit of the matter correlator~\rf{KMG}
sets in for $L\sim 1/\sqrt{\eps}$ to be
\bea
\lefteqn{C_c(x,y;\sqrt{u}) }\non & &
\propto \frac{2\sqrt{u}}{(x-y)^2+2u(x+y)xy+u^2x^2y^2}
\label{macrosolution}
\eea
with $u\propto L^2 \eps$.
The expression~\rf{macrosolution} obeys the following convolution property
\bea
\frac 1\pi\int_0^\infty dt\; {t}^{\frac 32} C_c(x,t;\sqrt{u}) C_c(t,y;\sqrt{v})
\non = C_c(x,y;\sqrt{u}+\sqrt{v})
\label{convolution}
\eea
which is analogous to that~\cite{DMSW93} for the Gaussian case.

\subsubsection{$\gamma_{str}=0$}

The continuum theory near the critical line $\a\approx -1$ looks
like 2d gravity + 1d matter.
The continuum spectral density reads
\be
\rho_c(\xi) \propto  \frac{1}{\pi} \sqrt{\xi^2-4\Lambda} .
\label{6.33}
\ee
The Itzykson--Zuber correlator
\be
C(\nu,\l;1) = \frac{4 \b^2}{4\b^2-1} +{\cal O}(\eps)
\label{6.34}
\ee
is non-singular if $\b\neq 1/2$. There is, hence, no unusual
behavior of the matter correlators in this case.

\subsubsection{Tricritical point}

At the tricritical point $\b=1/2$
$\a=-1$, the continuum system undergoes
a phase transition of the Kosterlitz--Thouless type
between the phases with $\gamma_{str}=-1/2$ and $\gamma_{str}=0$.

This domain is most interesting since
in the vicinity of the tricritical point the singular part
of the susceptibility
\be
\chi_0 = - \log{\left(\frac{1-\ka}{1+3\ka}\right)} -
2D \log{(\sqrt{2}\delta \b)}
\label{singb}
\ee
is $D$-dependent. Here the deviation of the tricritical point is
parametrized by
\be
\delta \a  = (3\ka+1)(1-\ka) (\delta \b)^2 .
\label{lines}
\ee
This region is the only one where $\gamma_{str}$ might depend on $D$
but it was not investigated in~\cite{Mak95}.

\subsection{Large-$D$ limit \label{l.D.}}

The Itzykson--Zuber integral
\be
I[\phi_x,\phi_y] \equiv \int dU \e^{cN\tr{\phi_x U^\dagger \phi_y U }},
\ee
which enters the partition functions~\rf{spartition},
can easily be calculated as $c\ra 0$:
\bea
\lefteqn{\ln{I[\phi_x,\phi_y]} = c \tr{\phi_x}\tr{\phi_y} } \nonumber \\
& &+
\frac{c^2 N^2}{2} \left[\ntr{\phi_x^2}-\left(\ntr{\phi_x}\right)^2\right]\non
& &~~~~
\times\left[\ntr{\phi_y^2}-\left(\ntr{\phi_y}\right)^2\right] +{\cal O}(c^3) .
\label{IZ}
\eea

If $V(\phi_x) \sim 1$ as $D\ra\infty$,
then $c \sim {1}/{D}$ for the kinetic term to be of order one and
only the first term is left on the RHS of \eq{IZ}.
The partition function~\rf{spartition} can be written as
\bea
\lefteqn{Z = \int \prod_x d\phi_x } \non
& &\times \e^{-N\sum_{x}\tr{V(\phi_x)}+ c \sum_{\{x,y\}}
\tr{\phi_x} \tr{\phi_y}} .
\label{exponent}
\eea

Further simplification occurs in the large-$N$ limit when we
can replace one trace
in the product of two traces in the exponent in~\rf{exponent} by the
average value due to factorization. One arrives, hence, at the one-matrix model
whose potential $\tV(\phi)$ is determined self-consistently from the equation
\be
\tV(\phi) =  V(\phi) - 2c D \LA \ntr{\phi}\RA_{\tilde{V}} \phi ,
\label{self-consistent}
\ee
where $\LA~\RA_{\tilde{V}}$ stands for the averaging in the one-matrix
model with the potential $\tilde{V}$.

For the one-cut solution of the Hermitean one-matrix model,
one rewrites \eq{self-consistent} as
\bea
\lefteqn{\tV^\p(\l) = V^\p(\l) -2 c D \left[
\int_{C_1} \frac{d\om}{4\pi i} \tV'(\om) \right.}
\non  & & \times\left.{\sqrt{(\om-x_-)(\om-x_+)}}
+ \frac{x_-+x_+}{2}  \right]
\label{eqfortV}
\eea
which is an equation for $\tV$.

Let us identify the cosmological constant with $g_1$ --- the coupling in front
of the linear term of the potential. For the susceptibility in genus zero one
gets
\be
\chi_0 = \ci \dot{V}(\om)\dot{E}_\om = \dot{\tilde{g}}_1 \frac{(x_--x_+)^2}{16}
\label{ch}
\ee
where $\dot{f}\equiv \partial f/ \partial g_1$, while \eq{eqfortV} yields
\be
\dot{\tilde{g}}_1 = \frac{1}{1+c D\frac{(x_--x_+)^2}{8}} .
\label{dottg}
\ee

To obtain the critical behavior, we expand near $x_-=x_{-}^{c}$ which
gives for a $k$-th multicritical point of the one-matrix model:
\be
x_--x_-^c \sim (\tilde{g}_1^c-\tilde{g}_1)^{1/k} .
\label{kth}
\ee
Under normal circumstances when \eq{4.14} holds,
one gets from~\rf{kth}
\be
\chi_0 -\chi_0^c \sim (\tilde{g}_1^c-\tilde{g}_1)^{1/k}
\ee
so that $\gamma_{str}=-1/k$
since
\be
(g_1^c-g_1)\sim (\tilde{g}_1^c-\tilde{g}_1) .
\ee

This is {\it not\/} the case, however, for
\be
c= - \frac{8}{D(x_--x_+)^2}
\label{ccric}
\ee
when the denominator in \eq{dottg} vanishes. At this point one has
\be
\dot{\tilde{g}}_1 \sim (x_--x_-^c)^{-1}
\ee
so that
\be
(g_1^c-g_1)\sim (\tilde{g}_1^c-\tilde{g}_1)^{(k+1)/k}
\sim (x_--x_-^c)^{k+1}.
\ee
One gets, therefore, for the susceptibility
\be
\chi_0 \sim (x_--x_-^c)^{-1} \sim  (g_1^c-g_1)^{-1/(k+1)}
\ee
which is associated with $\gamma_{str}=1/(k+1)$.

The formula~\rf{self-consistent}, which describes the reduction of the KM model
to a one-matrix model at large $N$ in the large-$D$ limit,
explicitly holds for the potential~\rf{V} when the exact solution is
known at any $D$. As $D\ra\infty$ with $c\sim 1/D$,
the potential $\tV$ coincides with the Penner potential
so that all the calculations can be explicitly done~\cite{Mak95}.
The only possible scaling behavior is with $\gamma_{str}=0$ in
perfect agreement with the results of Subsect.~3.4. This seems to be
a $k\ra\infty$ limiting case of $\gamma_{str}=1/(k+1)$ which appears from the
critical behavior with $\gamma_{str}=-1/k$ of the one-matrix model with
the potential $\tV$.

Nonvanishing results for continuum correlators  can be obtained in the
large-$D$ limit only for those of operators living at the same lattice site
while the Itzykson--Zuber correlator for a contour of the length $L$
is suppressed as
\be
C(\nu,\l;L) \sim c^L \sim D^{-L} .
\ee
Therefore extended correlators vanish in the large-$D$ limit.

\section{The meander problem~\protect{\cite{MP95}} }

The meander problem is known to people working on Quantum Field Theory since
the Arnold's question to V.~Kazakov in the middle of the eighties.
The problem is to calculate combinatorial numbers associated with
the crossings of an infinite river (Meander) and a closed road
by $2n$ bridges.%
\footnote{See~\cite{DGG95} for an introduction to the subject.}
Neither the river nor the road intersects with itself.
These principle meander numbers, $M_n$, obviously describe the number of
different foldings of a closed strip of $2n$ stamps or of a closed
polymer chain.

One can consider also a generalized problem of the multi-component meander
numbers $M_n^{(k)}$ which are associated with $k$ closed loops of the road
so that $M_n\equiv M_n^{(1)}$. The results of a computer enumeration
of the meander numbers are presented in \cite{LZ93,DGG95} up to $n=12$.

\subsection{Hermitean matrix model for meanders}

Meanders can be described by the following Hermitean matrix model~\cite{KK}
\bea
{\cal F}_{N\times N}(c)= \frac 2{N^2}
\int \prod_{a=1}^m dW_a \e^{-\frac N2 \sum_{a=1}^m \tr{W_a^2}} \non
\cdot \ln{\left(\int d\phi \e^{-\frac N2 \tr{\phi^2}
+\frac {cN}2 \sum_{a=1}^m \tr{\left(\phi W_a \phi W_a\right)}}\right)}
\label{Hpartition}
\eea
where the integration goes over the $N\times N$ Hermitean matrices
$W_a$ ($a=1,\ldots,m$) and $\phi$. The logarithm in \eq{Hpartition}
leaves only one closed loop of the field $\phi$. The coupling constant
$c$ is associated with the (quartic) interaction between $W_a$ and $\phi$.

Expanding the generating function~\rf{Hpartition} in $c$ and identifying
the diagrams with the ones for the meanders, one relates the large-$N$ limit
of ${\cal F}_{N\times N}(c)$ with the following sum over the meander numbers
\bea
\lim_{N\ra\infty} {\cal F}_{N\times N}(c) =
 \sum_{n=1}^\infty \frac {c^{2n}}{2n} \sum _{k=1}^n M_n^{(k)} m^k .
\label{Zm}
\eea
The $N\ra\infty$ limit is needed to keep only planar diagrams as in the meander
problem.

The RHS of \eq{Hpartition} can be expressed entirely via the Gaussian
averages of $W$'s. This leads to the following representation
of the meander numbers:
\bea
\lefteqn{\sum _{k=1}^n M_n^{(k)} m^k =
\sum_{a_1,a_2,\cdots, a_{2n-1}, a_{2n}=1}^m} \non & &
\times \LA \frac 1N \tr{W_{a_1} W_{a_2} \cdots W_{a_{2n-1}} W_{a_{2n}}}
\RA^2_{\hbox{\footnotesize{Gauss}}}
\label{words}
\eea
where the average over $W$'s is calculated with the Gaussian weight ---
the same as in~\rf{Hpartition}.
This formula can be proven by calculating the Gaussian integral over $\phi$
in \eq{Hpartition}, expanding the result in $c$ and comparing with
the RHS of \eq{Zm}. The factorization at large $N$ is also used.

The principle meander numbers $M_n$ are given by~\eq{words}
as the linear-in-$m$-terms, \ie as linear terms of the expansion in $m$.
This looks like the replica trick
which suppresses higher loops of the field $W$.

The ordered but cyclic-symmetric sequence of indices
$a_1,a_2,\ldots, a_{2n-1}, a_{2n}$ is often called as a {\em word\/}
constructed
of $m$ letters. The average on the RHS of \eq{words} is the meaning
of a word. Thus, the meander problem in equivalent to summing
over all the words with the Gaussian meaning.

Since for $m=1$
\be
\LA \frac 1N \tr{W^{2n}}\RA_{\hbox{\footnotesize{Gauss}}}=
\frac{(2n)!}{(n+1)!n!}\equiv C_n ,
\label{Catalan}
\ee
which is known as the Catalan number of the order $n$, one gets from
\eq{words}
\be
\sum _{k=1}^n M_n^{(k)} = C_n^2 .
\label{1st}
\ee
This is nothing but the first sum rule of~\cite{DGG95}.

\subsection{Complex matrix model for meanders}

The meander numbers can
alternatively be represented as the Gaussian average over the complex
matrices. The corresponding generating function reads
\be
{\cal F}(c)= \frac 1{N^2}
\LA \ln{\left(\int d\phi_1 d\phi_2 \e^{-S}\right)}
\RA_{\hbox{\footnotesize{Gauss}}}
\label{Cpartition}
\ee
with
\bea
\lefteqn{S=\frac N2 \tr{\phi^2_1}+\frac N2 \tr{\phi^2_2} } \non
& &-cN \sum_{a=1}^m \tr{\left(\phi_1 W^\dagger_a \phi_2 W_a\right)} .
\label{action}
\eea
Here, $\phi_1$ and $\phi_2$ are Hermitean while $W_a$ ($a=1,\ldots,m$)
are general complex matrices.

It is convenient to introduce one more generating function
\bea
\lefteqn{M(c) = c }\non & &\times \LA \frac{\int d\phi_1 d\phi_2 \e^{-S}
\frac 1N \tr{\phi_1 W^\dagger_1 \phi_2 W_1}}
{\int d\phi_1 d\phi_2 \e^{-S}} \RAG
\label{defM}
\eea
where only one component of $W_a$, say the first one, enters the averaging
expression. Differentiating the generating function~\rf{Cpartition}
with respect to $c$ and noting that all $m$ components of $W_a$ are
on equal footing, we get the relation
\be
c\frac{d{\cal F}(c)}{dc} = m M(c)
\ee
between the two generating functions.

In order to show how the complex matrix model recovers
the meander numbers, let us
 replace $\phi^{ij}_1$ or $\phi^{ij}_2$ in the numerator of \eq{defM} by
 $N^{-1}\partial/\partial \phi^{ji}_1$ or
$N^{-1}\partial/\partial \phi^{ji}_2$, respectively, and integrate by parts.
Repeating this procedure iteratively, we get
\be
M(c)= \sum_{n=1}^\infty c^{2n}
\sum _{k=1}^n M_n^{(k)} m^{k-1}  ,
\label{Cm}
\ee
with
\bea
\lefteqn{\sum _{k=1}^n M_n^{(k)} m^{k-1}
= \sum_{a_2,\cdots, a_{2n-1}, a_{2n}=1}^m } \non & &
 \times \LA \frac 1N \tr{W_{1} W^\dagger_{a_2} \cdots
W_{a_{2n-1}} W^\dagger_{a_{2n}}}
\RA^2_{\hbox{\footnotesize{Gauss}}}\hspace*{-.5cm} .
\label{Cwords}
\eea
Equation~\rf{Cwords} can alternatively be derived by calculating
the Gaussian integrals over $\phi_1$ and $\phi_2$ in \eq{defM} by
virtue of
\bea
\lefteqn{\int d\phi_1 d\phi_2 \e^{-S}
=\det{}^{-1/2}  \Big[ {\rm I}\otimes {\rm I} }\non & &  -c^2
\sum_{a,b=1}^m  W_a W^\dagger_b \otimes
\left(W_b W^\dagger_a \right)^{\rm t }\Big].
\label{cdeterminant}
\eea

For $m=1$ the formula
\be
\LA \frac 1N \tr{(W W^\dagger)^{n}}\RAG= C_n ,
\label{cCatalan}
\ee
which is analogous to \eq{Catalan}, holds for the complex matrices.
This results again in \eq{1st}.

It is instructive to consider also the case when $W$ is a fermionic Grassmann
valued matrix {\it \'a la}\/ \cite{MZ94}.%
\footnote{See~\cite{SS95} for a review.}
We shall denote the fermionic
matrix as $F$ and its conjugate as $\bar{F}$. Then we get~\cite{MZ94}
\bea
\lefteqn{\LA \frac 1N \tr{\left(F \bar{F}\right)^{n}} \RAG }\non & &=
\left\{
\begin{array}{cl}
0 & n=2p ~\hbox{(even)} \\
C_p & n=2p+1 ~\hbox{(odd)}
\end{array}
\right. .
\label{fCatalan}
\eea
Since each loop of the fermionic field is accompanied by a factor of $(-1)$,
we arrive at the sum rule
\bea
\lefteqn{\sum _{k=1}^n  (-)^{k-1} M_n^{(k)} } \non & &=
\left\{
\begin{array}{cl}
0 & n=2p ~\hbox{(even)} \\
C^2_p & n=2p+1 ~\hbox{(odd)}
\end{array}
\right. .
\label{2nd}
\eea
This is nothing but the second sum rule of~\cite{DGG95}.

Note that the trace of the square of a fermionic matrix vanishes
because of the anticommutation relation imposed on the components.
This is why we did not consider Hermitean fermionic matrices and
used first a representation
of meanders in terms of complex matrices to discuss fermionic representation
of meanders. Fermionic matrix models are a natural representation of
the notion of the signature of arch configurations of~\cite{DGG95}.

\subsection{Supersymmetric matrix model for principle meander}

Having the representation~\rf{Cm}, \rf{Cwords} of meanders via general
complex matrices (either bosonic or fermionic), we can utilize the
idea of supersymmetry to kill the loops of the $W$-field instead
of the replica trick. Let us consider the two-component $W_a$
whose first component is bosonic while the second one is a fermionic matrix:
\bea
W_a= \left( B,F\right),~~~ \bar{W}_a \equiv
W^\dagger_a= \left( B^\dagger,\bar{F}\right).
\label{contentW}
\eea

Then all the multi-component meanders in Eqs.~\rf{Cm} or \rf{Cwords}
vanish and we get the following representation for the principle
meander
\bea
\lefteqn{ M_n = \sum_{a_2,\cdots, a_{2n-1}, a_{2n}=1}^2 } \non
& &\times \LA \frac 1N \tr{B \bar{W}_{a_2} \cdots
W_{a_{2n-1}} \bar{W}_{a_{2n}}}
\RAG \non & &\times
\LA \frac 1N \tr{\bar{W}_{a_{2n}}
W_{a_{2n-1}} \cdots  \bar{W}_{a_{2}} B}\RAG
\label{Swords}
\eea
where we kept trace of the order of matrices of how it appears
from \eq{Cm}. The signs are essential for fermionic components.

The generating function~\rf{Cpartition} equals zero for the
supersymmetric model since all the loops of the $B$ and $F$ fields
are mutually cancelled. One should use alternatively the generating
function~\rf{defM} which can be represented for the supersymmetric
matrix model as
\bea
\lefteqn{M(c) =} \non & & \LA \frac 1N \tr{B B^\dagger}
\ln{\left(\int d\phi_1 d\phi_2 \e^{-S}\right)} \RAG
\label{correlator}
\eea
where $S$ is explicitly given by
\bea
\lefteqn{S=\frac N2 \tr{\phi^2_1}+\frac N2 \tr{\phi^2_2}
-cN \tr{\left(\phi_1 B^\dagger \phi_2 B\right)} } \non & &
-cN \tr{\left(\phi_1 \bar F \phi_2 F\right)}
\label{Saction}
\eea
as is prescribed by \eq{action} with $W_a$ substituted
according to \eq{contentW}.
The equivalence of Eqs.~\rf{defM}
and \rf{correlator} in the supersymmetric case can be proven
replacing $B$ in the integrand on the RHS of \eq{correlator}
by $N^{-1}\partial/\partial B^\dagger$, integrating by parts,
and recalling that ${\cal F}(c)=0$.

Equation~\rf{Swords} is a nice representation of
the principle meander which looks more natural than the one based on the
replica trick. A hope is that it will be simpler to solve the $m=2$
supersymmetric model than a pure bosonic one at arbitrary $m$.

The total number of nonvanishing terms on the RHS of \eq{Cwords}
for the pure bosonic case,
which we shall denote as $\#_n$, is given by the following generating
function
\be
\sum_{n=0}^\infty \#_n c^{2n} =
\frac{\frac m2 \sqrt{1-4(m-1)c^2}-\frac m2 +1}{1-c^2m^2} .
\label{numgraphs}
\ee
This formula can be derived~\cite{MP95} using
non-commutative free random variables.

\subsection{Relation to the KM model}

Equation~\rf{numgraphs} is known from the solution~\cite{Gro92} of the KM
model with the Gaussian potential. There is the following
reason for that. Suppose that the matrices $W_a$ are unitary instead of
the general complex ones. Then one has
\bea
\lefteqn{\LA \frac 1N \tr{U_{a_1} U^\dagger_{a_2} \cdots
U_{a_{2n-1}} U^\dagger_{a_{2n}}} \RAH} \non & &=
\left\{
\begin{array}{cl}
1 & \hbox{for closed loops} \\
0 & \hbox{for open loops}
\end{array}
\right.
\label{meaning}
\eea
where the average over the unitary matrices $U$'s is over the Haar
measure.

Here the loops represent the sequences of indices
$\{a_1,a_2,\ldots, a_{2n-1}, a_{2n} \}$. The nonvanishing result is only
when the loop is closed and encloses a surface of the vanishing minimal area,
\ie each link of the loop is passed at least twice.
This is a reflection of the so-called local confinement in the
KM model.

The generating function~\rf{numgraphs} coincides with the following
correlator in the KM model with the Gaussian potential
on an infinite $D$-dimensional lattice
\be
\sum_{n=0}^\infty \#_n c^{2n} =
\LA \frac 1N \tr{\phi^2(0)} \RA  ,
\label{KMcorrelator}
\ee
where the average is defined with the same weight as in \eq{spartition},
provided that $m=2D$.

The solution of the
KM model with the Gaussian potential can be completely
reformulated as a combinatorial problem of summing over
all such closed loops of a given length with all possible backtrackings
(or foldings) included. Its solution~\cite{Mak92} is given by \eq{numgraphs}.

By virtue of the Eguchi--Kawai reduction~\cite{EK82}%
\footnote{See~\cite{Das87} for a review.},
the correlator~\rf{KMcorrelator} in
the KM model on the infinite lattice is equivalent
to that in the reduced model given by
\bea
\lefteqn{\sum_{n=0}^\infty \#_n c^{2n} = }\non & & \frac{
\int d\phi_1 d\phi_2 \prod_{a=1}^m dU_a
 \e^{-S\left[\phi,U\right]}\frac 1N \tr{\phi^2_1} }
{\int d\phi_1 d\phi_2 \prod_{a=1}^m  dU_a
 \e^{-S\left[\phi,U\right]} }
\label{KMrcorrelator}
\eea
with $m=2D$ and the reduced action being
\bea
\lefteqn{S\left[\phi,U\right] =
 \fr N2 \tr{\phi^2_1}  + \fr N2 \tr{\phi^2_2}} \non & & \hspace*{1.0cm}
-  c N \sum_{a=1}^m \tr{\left(\phi_1{U}^\dagger_a \phi_2
 U_a\right)} .
\label{KMraction}
\eea
The representation~\rf{KMrcorrelator}, \rf{KMraction} can now be rewritten
as
\bea
\lefteqn{\sum_{n=1}^\infty \#_n c^{2n}
 = c \sum_{a=1}^m}
 \non & & \hspace*{-.2cm}
 \times \LA \frac{\int d\phi_1 d\phi_2 \e^{-S\left[\phi,U\right]}
 \frac 1N  \tr{\phi_1 U^\dagger_a \phi_2 U_a}}
{\int d\phi_1 d\phi_2 \e^{-S\left[\phi,U\right]}} \RAH
\label{defrM}
\eea
since the determinant~\rf{cdeterminant} is equal to a constant when $W_a$
are unitary. The representation~\rf{defrM} looks very similar to
the generating function~\rf{defM} of the meander numbers. The difference
is that the average is over the unitary matrices in \eq{defrM} and
over the Gaussian complex matrices in \eq{defM}.

We can interpolate between the two cases by modifying the
weight for averaging over $W$'s along the line
of~\cite{IMS86}. Let us introduce
\bea
\lefteqn{\LA F\left[ W,W^\dagger \right]\RA_\alpha \equiv
\int \prod_{a=1}^m \Big(dW_a^\dagger dW_a  }
\non & & \hspace*{-.6cm} \times
\e^{-\fr{\alpha N}2\tr{\left({W}^\dagger_a W_a-1+\fr 1\alpha\right)^2}
+\fr N{2\alpha}} \Big) F\left[ W,W^\dagger \right] .
\eea
Then the averaging over the Gaussian complex matrices is
reproduced as $\alpha\ra0$ while the average over the unitary
matrices is recovered as $\alpha\ra\infty$ since
the matrix $W_\a$ is forced to be unitary as $\alpha\ra\infty$.

We see, thus, that the words are the same both for the meander problem
and for the KM model. The only difference resides in the meaning
of nonvanishing words --- it is equal to one for the unitary matrices.

\section{Conclusions}

\begin{itemize}\vspace{-6pt}
\addtolength{\itemsep}{-6pt}
\item
The Kazakov--Migdal--Penner model is an explicit example of crumpled strings
(= discretized random surfaces) in the $D>1$ embedding space.
It possesses the only continuum limits associated with lower
dimensional theories of $c=0$ or $c=1$.
\item
The meander problem results in a more complicated matrix model than those
solved before. It belongs to the same generic class of problems
of words as the  large-$N$ QCD in $D=4$ but is presumably simpler.
\item
The relation of the matrix models describing meanders with the
KM model might be a hint on how to solve the former ones.
\item
The supersymmetric matrix models of the type discussed
in connection with the meander problem could be useful for
discretization of super-Riemann surfaces and superstrings.
\end{itemize}

\end{document}